\documentclass[a4,12pt]{article}
\newcommand{\be}{\begin{eqnarray}}
\newcommand{\ee}{\end{eqnarray}}
\newcommand{\del}{\partial}
\makeatletter
\newcommand{\fslash}[2][0mu]{%
    \mathchoice
     {\fsl@sh\displaystyle{#1}{#2}}%
     {\fsl@sh\textstyle{#1}{#2}}%
     {\fsl@sh\scriptstyle{#1}{#2}}%
      {\fsl@sh\scriptscriptstyle{#1}{#2}}}
    \newcommand{\fsl@sh}[3]{%
    \m@th\ooalign{$\hfil#1\mkern#2/\hfil$\crcr$#1#3$}}
\def\lsim{\raise0.3ex\hbox{$<$\kern-0.75em\raise-1.1ex\hbox{$\sim$}}}
\def\gsim{\raise0.3ex\hbox{$>$\kern-0.75em\raise-1.1ex\hbox{$\sim$}}}
\makeatother
\usepackage{color}
\usepackage{epsfig}
\usepackage{graphics}

\title{
\bf Probing freeze-out conditions
in heavy ion collisions with
moments of charge fluctuations 
   }
  \author{ \\ F. Karsch$^{1,2}$
  and K. Redlich$^{3,4}$
    \\ \\
    \small
 $^1$ Fakult\"at f\"ur Physik, Universit\"at Bielefeld,\\
\small Postfach 100 131, D-33501 Bielefeld, Germany\\
\small  $^2$ Physics Department, Brookhaven National Laboratory, Upton, NY 31973, USA \\
  \small     $^3$ Institute of Theoretical Physics University of Wroclaw,\\
\small   PL-50204 Wroclaw, Poland\\
$^4$\small Theory Division, CERN, CH-1211 Geneva 23, Switzerland }
\date{July 15, 2010}

\begin{document}
\maketitle

\centerline{Abstract} {\small
We calculate the first four  moments of baryon number, electric charge and strange\-ness
fluctuations within the hadron resonance gas model. Different moments and their
ratios as well as skewness and kurtosis are evaluated on the phenomenologically
determined freeze-out curve in the temperature, baryon chemical potential
plane. The model results and its predictions as well as  relations between different
moments are compared with the first  data on net proton fluctuations in Au-Au collisions
obtained at RHIC by the STAR Collaboration. We find good agreement between the
model calculations and experimental results. We also point out that higher order
moments should be more sensitive to critical behavior and will also distinguish
hadron resonance gas model calculations from results obtained from lattice QCD.

\vspace{0.5cm}
 \mbox{} \hfill CERN-PH-TH-2010-161\\
\newpage
\section{Introduction}

One of the major goals in theoretical and experimental studies
of the thermodynamics of strongly interacting matter is to explore
the QCD phase diagram at non-zero temperature ($T$) and non-zero
baryon chemical potential ($\mu_B$) \cite{koch}. A central target of interest
is the 'critical point' - a second order phase transition point,
that has been postulated to exist in the $T$-$\mu_B$ phase diagram
\cite{CEP,CP}.
Although its existence is not yet firmly established, its presence would
result in large correlation lengths, {\it i.e.} large fluctuations
in various thermodynamic quantities. Remnants of these large
fluctuations may become accessible in heavy ion collisions through
an event-by-event analysis of fluctuations \cite{StephanovPP} in various
channels of
hadron quantum numbers (charges), eg. baryon number ($B$), electric charge ($Q$)
and strangeness ($S$). In fact, at vanishing baryon chemical
potential it has been shown that moments of charge fluctuations
are sensitive indicators for the occurrence of a transition from hadronic
to quark-gluon matter \cite{EKR}. They drastically change during the transition.

When comparing theoretical equilibrium calculations of charge fluctuations
with experimental results from heavy ion collisions,
the crucial question is, whether at the time of hadronization the thermal
system generated in these collisions has kept memory of the plasma
phase or the expansion period during which it may have passed by a critical point
\cite{rajagopal}. If so, this may lead to an enhancement of fluctuations
over ordinary thermal effects at freeze-out. On the other hand, it is well
known that basic features of hadronization in heavy ion collisions are well
described in terms of the hadron resonance gas model (HRG) \cite{stachel,andro1}.
The analysis of experimental data on the production cross sections
of various hadrons in heavy ion collisions shows astonishingly
good agreement with corresponding thermal abundances calculated
in a HRG model at appropriately chosen
temperature and chemical potential \cite{redlich,andronic}. Abundances
of  strange and non-strange mesons and baryons are consistently
described by freeze-out temperatures and baryon chemical potentials that
are a function of collision energy only.

If some memory of large correlation lengths persists in the thermal medium
at time of freeze-out this must be reflected in higher moments of charge
distributions. These moments are more sensitive to large correlation lengths
and thus relax more slowly to their true equilibrium values at the time
of freeze-out \cite{Stephanov}. It thus is an interesting question to test whether
also more detailed information on the thermal distribution of hadron
species, predicted by the HRG model, can be confirmed experimentally.
In fact, one may consider HRG model results on moments of charge
fluctuations as a theoretical baseline prediction for the currently
ongoing low energy heavy ion runs at RHIC \cite{lowE} and future studies
of charge fluctuations at the LHC. Any deviations from HRG
model predictions would constitute evidence for new phenomena at
the time of hadronization that have not been seen in experiments so far.

Of course, eventually thermodynamics at freeze-out should be described
by thermal QCD, eg. through lattice calculations. A direct comparison
of experimental and HRG model results on higher moments of charge fluctuations with
lattice calculations in the hadronic phase is possible
\cite{Tawfik},
but is still difficult as so far most lattice calculations are performed with
staggered fermions on rather coarse lattices. They need to be performed closer
to the continuum limit to reproduce the correct hadron spectrum \cite{bazavov,fodor}.
Fortunately the calculation of ratios of moments is less sensitive to such cut-off
effects \cite{EKR,Tawfik,Gavai}. At present, it seems that lattice QCD calculations
of ratios of moments of charge fluctuations, performed at non-vanishing chemical
potential by using
a Taylor expansion of thermodynamic observables \cite{Gavai6,Taylor}, are in good
agreement with HRG model calculations for temperatures below the transition
temperature. We will consider  this issue in more detail in Section 5 and indicate
 when such  agreement may break down.

In the following we will discuss the dependence of higher moments of fluctuations of
baryon number, electric charge and strangeness on the collision energy.
In particular, we will calculate ratios of quartic and quadratic (kurtosis), cubic
and quadratic (skewness) as well as quadratic charge fluctuations normalized to
their mean value along a phenomenologically determined freeze-out curve in heavy ion
collisions. We also discuss correlations of different charges, eg.
the correlation of baryon number and electric charge normalized to the squared
baryon number fluctuations.

In the next section we will summarize basic results on the parametrization
of the freeze-out curve in heavy ion collisions. In section 3 we introduce
moments of charge fluctuations and discuss their calculation in the
HRG model.
In section 4 we  compare the HRG model results with recent data
on net proton number fluctuations in heavy ion collisions
at RHIC energies obtained by the STAR Collaboration \cite{STAR} and discuss
additional fluctuation observables that may be analyzed to further characterize
thermal conditions at freeze-out. Section 5 is devoted to a comparison of the HRG
model  and lattice QCD calculations.  We give our conclusions in section 6.

\section{Freeze-out conditions in heavy ion collisions}

Abundances of  strange and non-strange mesons and baryons produced
in heavy ion collisions in a wide range of collision energies are consistently
described by freeze-out temperatures and baryon chemical potentials that
are a function of collision energy only. In fact, the freeze-out curve $T(\mu_B)$
in the $T$-$\mu_B$ plane and the dependence of the baryon chemical potential
on the center of mass energy in nucleus-nucleus collisions can be
parametrized by simple functions \cite{cleymans}
\begin{equation}
T(\mu_B) = a - b\mu_B^2 -c \mu_B^4 \; ,
\label{TmuB}
\end{equation}
where $ a =  (0.166 \pm 0.002)$ GeV, $b = (0.139 \pm 0.016)$ GeV$^{-1}$,
$c = (0.053 \pm 0.021)$ GeV$^{-3}$ and
\begin{equation}
\mu_B(\sqrt{s_{NN}}) = \frac{d}{1 + e\sqrt{s_{NN}}} \; , \label{muB}
\end{equation}
with $d = (1.308\pm 0.028)$ GeV and  $e = (0.273 \pm 0.008)$ GeV$^{-1}$.
This parametrization agrees with the phenomenological freeze out condition
of fixed energy per particle of about 1 GeV \cite{1gev}.

\begin{table}
\begin{center}
\begin{tabular}{|c|l|l|}
\hline
X & $d$~[GeV] & $e$~[GeV$^{-1}$] \\
\hline
B & 1.308(28) & 0.273 (8) \\
S & 0.214 & 0.161 \\
Q & 0.0211 & 0.106 \\
\hline
\end{tabular}
\caption{Parametrization of chemical potentials $\mu_X$ along the freeze-out curve using
the ansatz given in Eq.~\protect\ref{muB}.}
\label{tab:mu_fit}
\end{center}
\end{table}

The  energy dependence of strange and electric charge chemical potentials
are  obtained from the HRG model by demanding strangeness neutrality and isospin
asymmetry in the initial state of Au-Au collisions.
They can be parameterized in the same way
as it has been done for the baryon chemical potential in Eq.~\ref{muB}. The corresponding
fit parameters are given in Table~\ref{tab:mu_fit} and are shown in the left hand part of Fig. \ref{fig:par}.
The right hand part of this figure shows the variation of baryon number density,
$n_B=\langle N_B\rangle/V$,
and electric charge density, $n_Q=\langle N_Q\rangle/V$,  on the freeze-out curve.
The  strangeness density, $n_S=\langle N_S\rangle/V$, vanishes due to the imposed
strangeness neutrality condition.
We note that the ratio of baryon to strangeness chemical potential on the
freeze-out curve shows only a weak dependence on the collision energy,
\begin{equation}
\frac{\mu_S}{\mu_B} \simeq 0.164 + 0.018 \sqrt{s_{NN}} \ ,
\label{BSratio}
\end{equation}
which is in agreement with findings on freeze-out conditions at RHIC \cite{STAR1}.

\begin{figure}
\begin{center}
\includegraphics[width=6.9cm]{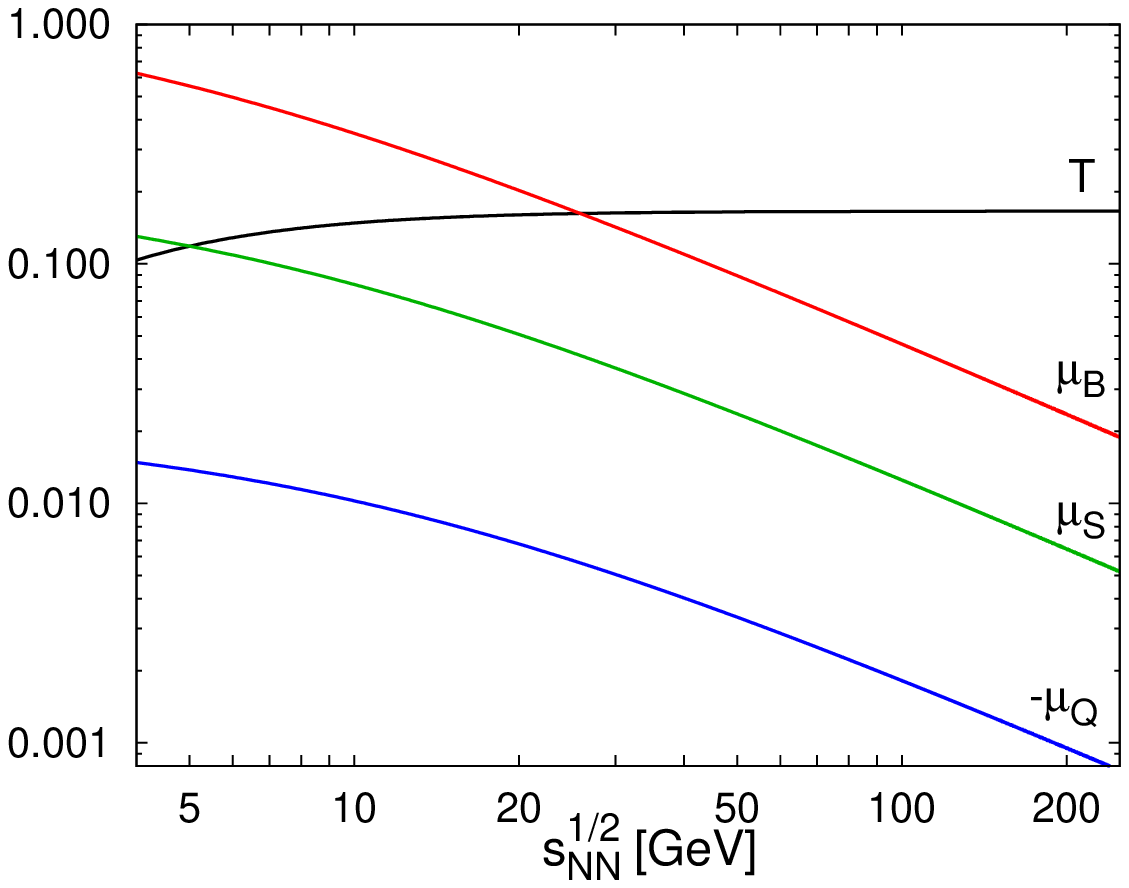}\hspace*{-0.5cm}
\includegraphics[width=6.9cm]{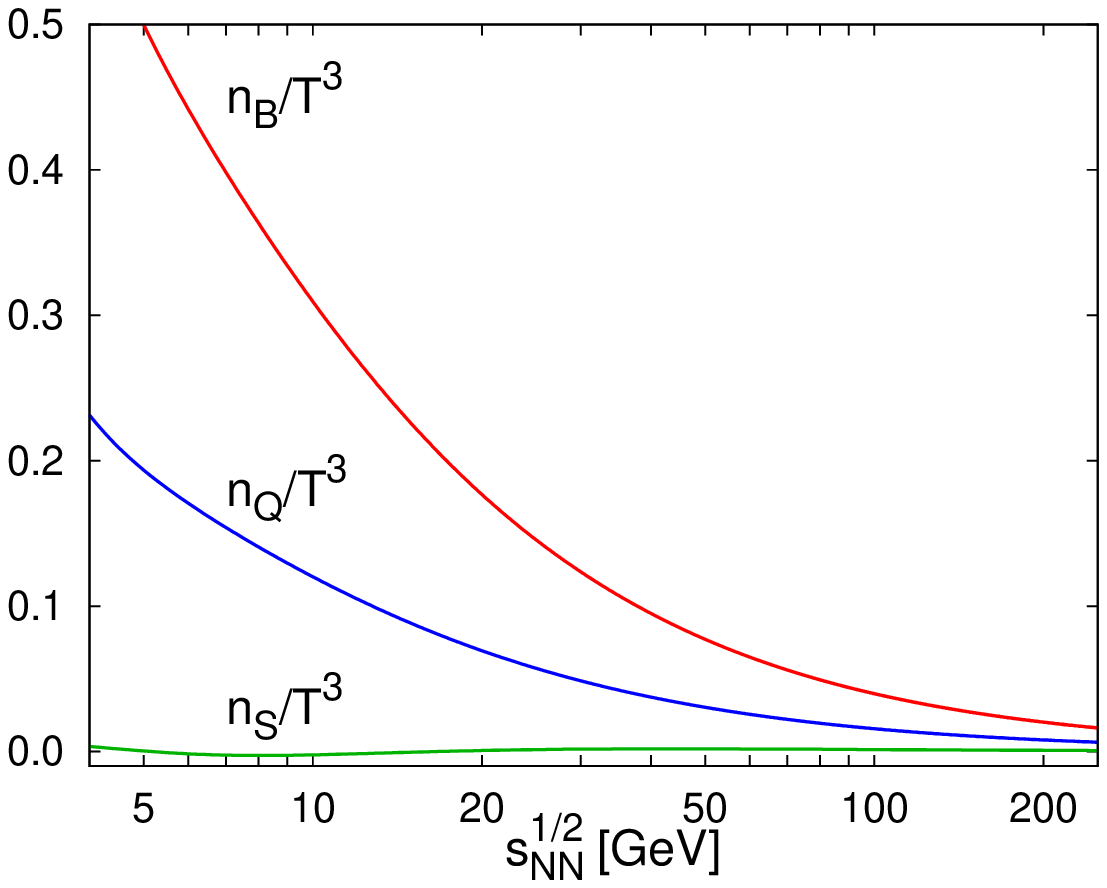}
\caption{
The energy dependence of temperature $T$ and chemical potentials for baryon number $\mu_B$, electric charge $\mu_Q$
and strangeness $\mu_S$ along the chemical freeze-out curve in units of GeV (left).
The right hand figure shows the variation of baryon number density $n_B$, electric charge density $n_Q$ and
strangeness density $n_S$ along the freeze-out curve.}
\label{fig:par}
\end{center}
\end{figure}

Current results from lattice calculations  \cite{bazavov,fodor,LGT_Tc} indicate that
at $\mu_B\simeq 0$
the transition temperature, within errors, coincides with the freeze-out temperature
extracted at the highest RHIC energy.
However, although firm statements from lattice QCD need to wait for calculations
closer to the continuum limit, it seems that for $\mu_B>0$
the curvature of the transition line $T(\mu_B)$ for $\mu_B>0$ is distinct from the
freeze-out curve \cite{LGT_crossover}.
This may indicate that at larger $\mu_B$ the QCD transition and the hadronic freeze-out
in heavy ion collisions are separated. The latter seems to occur at lower temperatures.

If this separation of transition and freeze-out curves is indeed confirmed and/or
hadronic fluctuations at freeze-out loose all memory of the passage through a possible
transition region, then not only particle yields but also charge fluctuations should be
characterized by thermal freeze-out conditions in the hadronic phase and may be
well described by the HRG model.

In the next section we will work out ratios of moments of various
conserved charges on the freeze-out curve using the HRG model.
In particular, we will calculate ratios of quadratic, cubic and
quartic fluctuations of baryon number.

\section{Fluctuations in the HRG model}

The basic quantity
that describes   thermodynamics at non-vanishing chemical
potential is the pressure. In the grand--canonical ensemble with baryon number, strangeness  and electric charge  conservation it is
obtained from the logarithm of the partition function as
\begin{equation} p(T,\mu_B,\mu_Q,\mu_S) = \lim_{V \rightarrow
\infty} {T\over V} \ln Z(T,\mu_B,\mu_Q,\mu_S, V) \quad . \label{pressure}
\end{equation}

In  the HRG model the partition function contains
all relevant degrees of freedom of the confined, strongly
interacting matter and implicitly includes interactions that
result in resonance formation.
The logarithm of the partition function in the HRG is obtained
as a sum over all stable hadrons and resonances and their anti-particles,
which can  be separated into
contributions from baryons and mesons,
\begin{equation}
\ln Z(T,\mu_B,\mu_Q,\mu_S)=\sum_{i\; \in\; {\rm mesons}} \ln
Z_{i}^+(T,\mu_Q,\mu_S) +
\sum_{i\; \in\; {\rm baryons}} \ln Z_{i}^-(T,\mu_B, \mu_Q,\mu_S) \quad
\label{eqq1}
\end{equation}
 The contribution of each  particle species  of mass $m_i$, spin degeneracy factor $g_i$,  carrying
baryon number $B_i$, electric charge $Q_i$ and  strangeness $S_i$ is expressed as
\begin{equation}
\ln Z_i^\pm (T,V,\vec\mu)={{VT}\over
{2\pi^2}}g_im_i^2\sum_{k=1}^\infty {{(\pm 1)^{k+1}}\over {k^2}}
 K_2({{k m_i/T}})\exp({k{{\vec c_i\vec\mu}/T}}) \ , \label{eqq2}
\end{equation}
where
$\vec{c}_i=(B_i,Q_i,S_i)$, $\vec{\mu}=(\mu_{B},\mu_{Q},\mu_{S})$
and $K_2(x)$ is the modified Bessel function.  The upper and lower signs are
for bosons and fermions, respectively. The first term in Eq.~\ref{eqq2} corresponds
to the Boltzmann approximation.

The partition function (Eq.~\ref{eqq1}) together with Eq.~\ref{eqq2} provides the basis
for the description of the thermodynamics of a system composed of hadrons and resonances
being in thermal and chemical equilibrium.  In the following we will focus on HRG model
predictions on fluctuations of conserved charges and their higher moments.

Fluctuations of the net charge  $N_q$ and its higher moments are obtained from
 derivatives of the logarithm of the partition function with respect to the
corresponding chemical potential. These derivatives define generalized 
susceptibilities\footnote{Note that we introduce here dimensionless susceptibilities.
This notation differs from that used in Ref.~\cite{STAR} by a factor $T^{n-4}$.}
\begin{equation}
\chi^{(n)}_q=\frac{\del^n[p\,(T,\vec\mu)/T^4]}{\del(\mu_q/T)^n} \; .
\label{eqq:6}
\end{equation}
The first derivative, $\chi^{(1)}_q$, is related to the mean value $M_q$ of the net
charge $N_q $,
 \begin{equation}
M_q \equiv \langle N_q \rangle =VT^3\chi^{(1)}_q \; ,
\label{eqq:7}
\end{equation}
{\it i.e.} it is the charge density in units of $T^3$, $\chi^{(1)}_q \equiv n_q/T^3$.
The second derivative with respect to $\mu_q/T$ gives the variance
$\sigma_q^2=\langle(\delta N_q)^2\rangle$ as
\begin{eqnarray}
\sigma_q^2 &=& VT^3\chi^{(2)}_q \; ,
\label{eqq:8}
\end{eqnarray}
with $\delta N_q= N_q-\langle N_q\rangle$.
The third $\chi^{(3)}_q$ and  the  fourth $\chi^{(4)}_q$ order moments can be expressed
in terms of $\delta N_q$ as
\begin{eqnarray}
 \langle(\delta N_q)^3\rangle & =& VT^3\chi^{(3)}_q \; ,
\label{eqq:9} \\
\langle(\delta N_q)^4\rangle-3\langle(\delta N_q)^2\rangle^2 &=& VT^3\chi^{(4)}_q \; .
\label{eqq:10}
\end{eqnarray}
We also introduce skewness ($S_q)$ and kurtosis ($\kappa_q$), which are generally used to characterize
the shape of statistical distributions,
\begin{eqnarray}
S_q\equiv  {{\langle(\delta N_q)^3\rangle}\over {\sigma_q^3}}\;\; ,\;\;
\kappa_q\equiv  {{\langle(\delta N_q)^4\rangle}\over {\sigma_q^4}}-3 \; .
\label{eqq:11}
\end{eqnarray}
Using Eqs.~\ref{eqq:7} to \ref{eqq:11}  we may relate  mean values,
fluctuations, skewness and kurtosis to the generalized susceptibilities
introduced in Eq.~\ref{eqq:6},
\begin{equation}
\frac{\sigma_q^2}{M_q} = \frac{\chi^{(2)}_q}{\chi^{(1)}_q} ,\;\; ~~~~
S_q\sigma_q = \frac{\chi^{(3)}_q}{\chi^{(2)}_q} ,\;\; ~~~~ \kappa_q \sigma_q^2 = \frac{\chi^{(4)}_q}{\chi^{(2)}_q}  \; .\;
\label{eqq:13}
\end{equation}
Finally we also consider correlations of charges, which can be obtained as mixed
derivatives of the pressure with respect to chemical potentials for charge $X$ and $Y$,
respectively,
\begin{equation}
\chi_{XY}^{(nm)} =  \frac{\partial^{n+m}p(T,\vec{\mu})/T^4}{\partial(\mu_X/T)^n
\partial(\mu_Y/T)^m}
 \; .
\label{obs}
\end{equation}
We will discuss the behavior of the ratios $\chi_{BS}^{(1m)}/\chi_B^{(2)}$  and
$\chi_{BQ}^{(1m)}/\chi_B^{(2)}$ for $m=1,\ 3$ on the freeze-out curve.

The above relations can be used to analyze properties of a thermodynamic system in equilibrium.
If the evolution of a heavy ion collision results in an equilibrated state that lost all its memory of
the previous evolution and hadronizes according to a hadronic statistical model, e.g.
the HRG model,
this will be reflected not only in particle yields but also in various moments of charge fluctuations.
As higher moments become increasingly sensitive to large correlation lengths,
any deviations from HRG model predictions at freeze-out could indicate  that
the system memorizes that it went through a region with large correlation lengths.

In the following we compare HRG model predictions with first measurements of the kurtosis, skewness  and variance
of net proton multiplicity distributions at mid-rapidity for Au+Au collisions at $\sqrt {s_{NN}} =$19.6,
62.4 and 200 GeV performed by the STAR collaboration \cite{STAR}.

\section{Charge fluctuations on the hadronic freeze-out curve}

In the Boltzmann approximation, which is a suitable approximation in the parameter range considered,
the HRG model provides a simple result for the thermodynamic pressure,
 \begin{equation}
{P\over {T^4}}={1\over {\pi^2}}\sum_i d_i ({{m_i/T}})^2K_2(m_i/T)
\cosh[( {B_i\mu_B+S_i\mu_S+Q_i\mu_Q})/T] \; .
\label{eqq:14}
\end{equation}
The contribution of anti-particles is explicitly included in Eq.~\ref{eqq:14}
through the $\cosh[x]$-term. Thus, the summation  is to be taken only over stable hadrons
and resonances.

A specific dependence of the pressure given in Eq.~\ref{eqq:14} on chemical potentials
implies definite predictions for ratios of different moments of charge fluctuations.
In particular, from
Eqs.~\ref{eqq:6} and \ref{eqq:14},  it is immediately clear, that ratios of fourth ($\chi^{(4)}_B$) and
second ($\chi^{(2)}_B$) order moments of the baryon number fluctuation
as well as the ratio of third ($\chi^{(3)}_B$) and first order moments are unity,
\begin{equation}
 \frac{\chi^{(4)}_B}{\chi^{(2)}_B}=1 ,\;\; ~~~~  \frac{\chi^{(3)}_B}{\chi^{(1)}_B}=1  \; , \;
\label{eqq:15}
\end{equation}
irrespective of the values of chemical potentials and temperature.
A direct consequence of the above HRG model results is that simple relations hold for the
mean net baryon number, kurtosis and skewness,
\begin{equation}
\kappa_B \sigma_B^2  = 1 ~~\; ,\;~~
\kappa_B  M_B = S_B\sigma_B \; .
\label{eqq:16}
\end{equation}
The first relation is straightforward  from Eq.~\ref{eqq:13}, the second is valid since
\begin{equation}
\kappa_B M_B= \frac{\chi^{(4)}_B}{\chi^{(2)}_B}{{M_B}\over{\sigma^2}} = \frac{\chi^{(4)}_B}{\chi^{(2)}_B}
\frac{\chi^{(1)}_B}{\chi^{(2)}_B} =
\frac{\chi^{(4)}_B}{\chi^{(2)}_B}
\frac{\chi^{(1)}_B}{\chi^{(3)}_B}
\frac{\chi^{(3)}_B}{\chi^{(2)}_B}=
 S_B \sigma_B.
 \label{eqq:17}
\end{equation}
In the HRG model the skewness  $S_B\sigma_B$  can be explicitly obtained from
\begin{equation}
 S_B \sigma_B={{\sum_{i\; \in\; {\rm baryons}} d_i ({{m_i/T}})^2K_2(m_i/T)
\sinh[( {\mu_B+S_i\mu_S+Q_i\mu_Q})/T]}\over
 {\sum_{i\; \in\; {\rm baryons}} d_i ({{m_i}/T})^2K_2(m_i/T)
\cosh[( {\mu_B+S_i\mu_S+Q_i\mu_Q})/T]}} \; .
\label{eqq:18}
\end{equation}
Consequently, for $\mu_S=\mu_Q=0$ one  gets
\begin{equation}
S_B\sigma_B=  \tanh(\mu_B/T).
\label{eqq:19}
\end{equation}
This simple result arises from the fact that in the HRG model only
baryons with baryon number $B=1$ contribute to the various moments.
\begin{figure}
\begin{center}
\epsfig{file=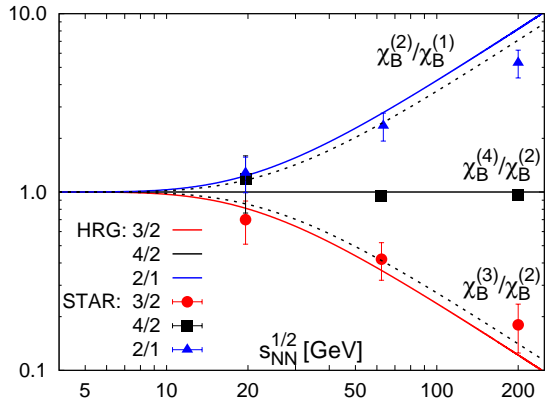,width=8.0cm}
\end{center}
\caption{\label{skew}
The ratio of quadratic fluctuations and mean net baryon number ($\sigma_B^2/M_B$),
cubic to quadratic  ($S_B\sigma_B$) and
quartic to quadratic ($\kappa_B\sigma_B^2$) baryon number fluctuations
calculated in the HRG model on the freeze-out curve and compared to results
obtained by the STAR collaboration \cite{STAR}.
The dashed curves show the approximate $\tanh (\mu_B/T)$
result for $\kappa_B\sigma_B^2$ and $S_B\sigma$, respectively.}
\end{figure}

In heavy ion collisions  the strangeness  and electric charge chemical potentials are much smaller
than $\mu_B$ (see Fig. 1). The above relation thus can be considered to be a good estimate of skewness
at chemical decoupling.
We will show in the following that corrections due to non-vanishing electric charge
and strangeness chemical potentials are indeed small for baryon number fluctuations.

\subsection{Comparison of the HRG model results on baryon number fluctuations with RHIC data}

The relations for skewness and kurtosis summarized in Eqs.~(\ref{eqq:16}), (\ref{eqq:18}) and (\ref{eqq:19})
are generic results, expected to hold if thermodynamics is governed by the HRG model.
Knowing the energy dependence of thermal parameters along the freeze-out curve
(Eqs.~\ref{TmuB} and \ref{muB})
we can directly verify if these particular relations, deduced within the HRG model,
are consistent with
recent findings of the STAR collaboration, which measured moments of baryon number fluctuations through
net-proton number fluctuations \cite{STAR}.

Fig.~\ref{skew} shows a comparison of the energy dependence of quadratic fluctuations
($\sigma_B^2$) normalized to the net baryon number ($M_B$),  skewness $S_B\sigma_B$ and
kurtosis $\kappa_B\sigma_B^2$ obtained in the HRG model at chemical freeze-out with the
STAR data.

Obviously, the HRG model provides a good description of properties of different
moments of net proton number fluctuations measured at RHIC energies. The reason for
considering ratios of charge fluctuations rather than absolute values for different
moments was, of course, that one is independent of definitions of the interaction
volume and also is less sensitive to experimental cuts and systematic errors. 
Moreover, some of these
ratios have an interesting  interpretation, like  e.g. the ratio
$\chi_B^{(4)}/\chi_B^{(2)}$ which directly characterizes the dominant degrees
of freedom carrying baryon number \cite{EKR}.}

In addition it is, of course, also of interest to understand whether the
HRG model can quantitatively describe the energy dependence of the STAR
data  \cite{STAR} on the first four moments, i.e.  the mean, variance, skewness and
kurtosis\footnote{Of course, as we have already verified consistency of three
ratios with the HRG model calculations, only the energy dependence of one of theses
observable provides additional information.}.

In order to compare the HRG model calculations with the experimental results
presented in \cite{STAR} we note that this analysis only explored fluctuations
in a limited phase space. In fact, the data on mean particle yields
differ from  previous results obtained by the STAR Collaboration 
\cite{STAR1}. From Ref. \cite{STAR}  one  gets: $M_{p-\bar{p}}\simeq 1.75\pm 0.25$
and $M_{p-\bar{p}}\simeq 3.5\pm 0.4$ in the central (0-5$\%$) bin of Au-Au collisions at
$\sqrt {s_{NN}} =200$ GeV and  62.4 GeV, respectively. These values should be compared
with $M_{p-\bar{p}}\simeq 8\pm 1.8$ and  $M_{p-\bar{p}}\simeq 15.4\pm 2.1$  obtained
at mid-rapidity  at corresponding energies in \cite{STAR1}.
These data differ by a common factor $K\simeq 0.22$.
Part of the difference may be attributed to the fact that net proton
fluctuation data in Ref. \cite{STAR} were taken
in the restricted transverse momentum range  $0.4 < p_T < 0.8$ GeV.

In the HRG model used by us the thermal phase--space of all particles is not
restricted. Consequently, in order to compare predictions for different moments of
net proton fluctuations with experimental data one needs to rescale its thermal
phase space by the above mentioned factor $K\simeq 0.22$. Effectively, this
corresponds to rescaling  the volume parameter $VT^3$ appearing for instance
in Eq.~\ref{eqq:7}, by the  $K$-factor,
although its origin is not necessarily related with a smaller volume of the system
at chemical freeze out. 
\begin{figure}
\begin{center}
\epsfig{file=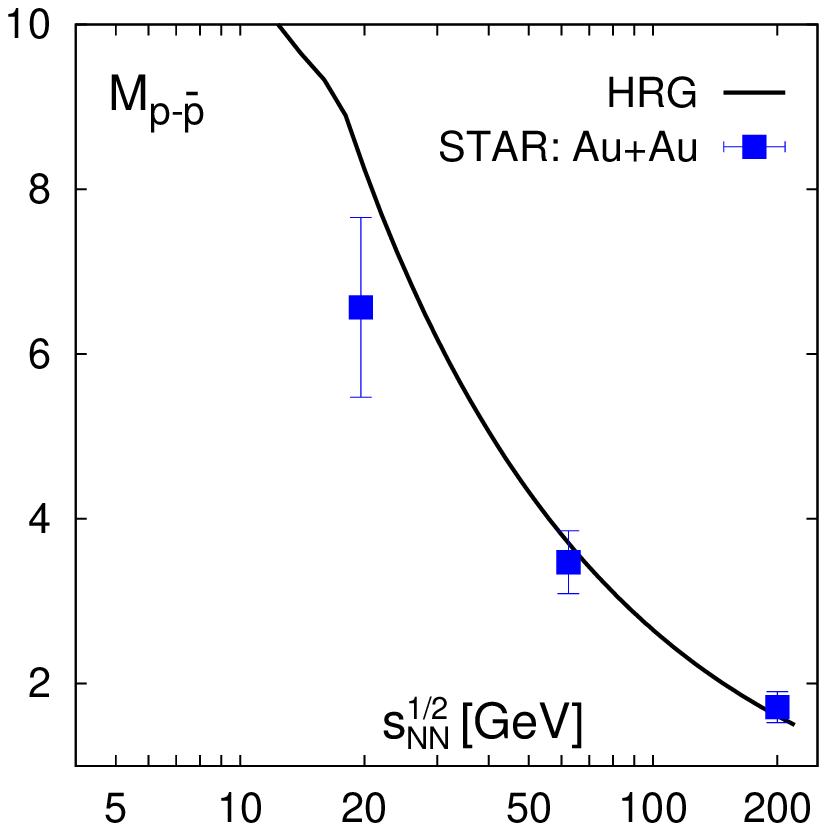,width=7.5cm}\hspace*{-1.5cm}
\epsfig{file=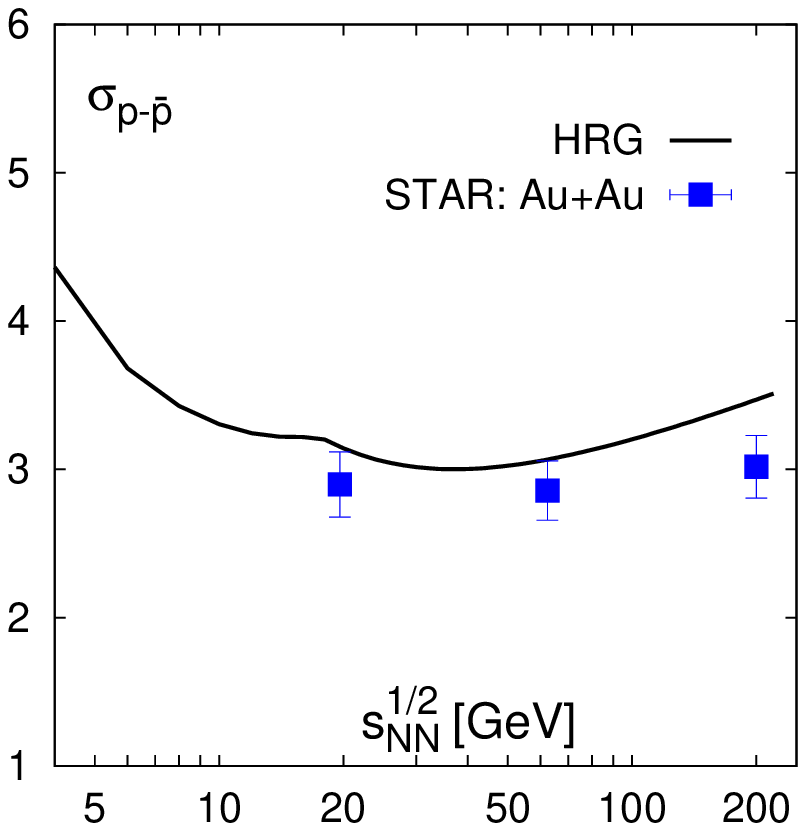,width=7.5cm}
\vspace*{-0.5cm}
\epsfig{file=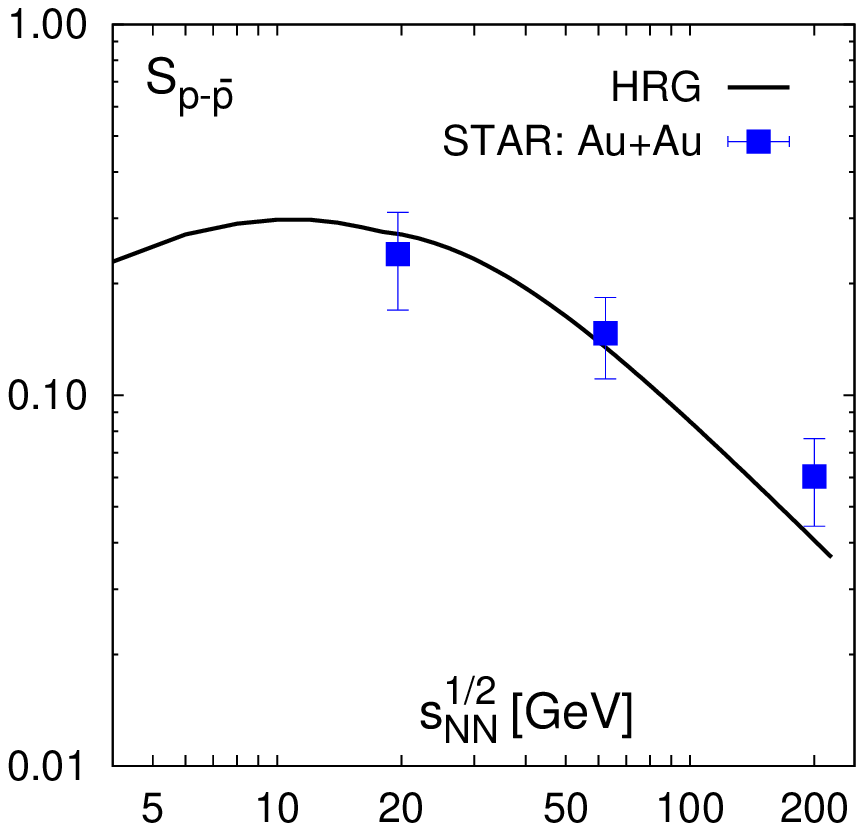,width=7.5cm}\hspace*{-1.5cm}
\epsfig{file=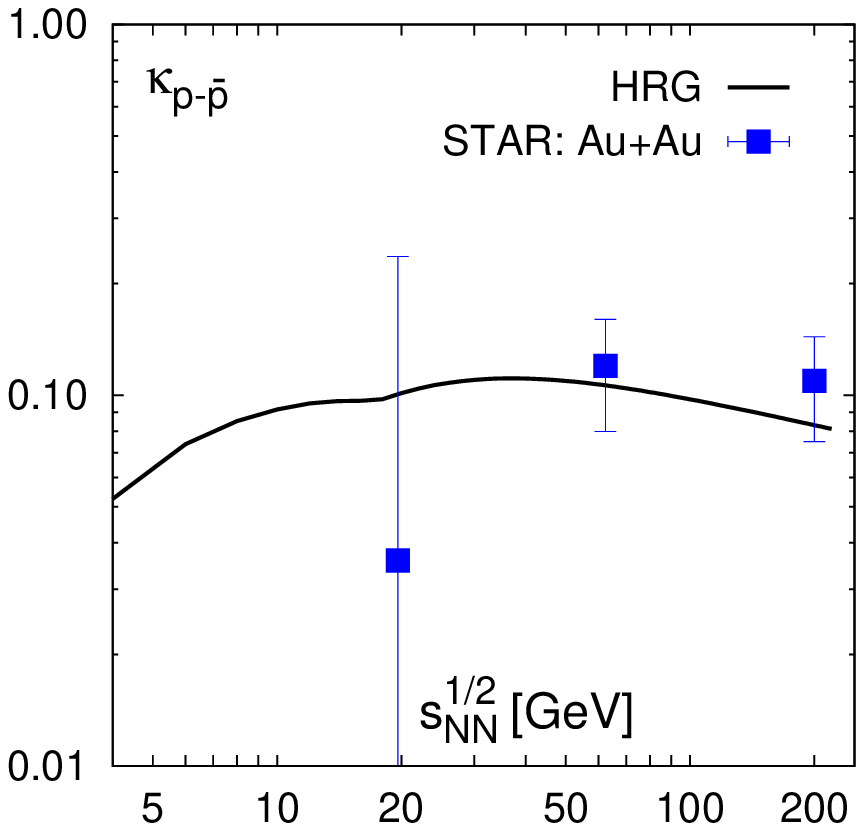,width=7.5cm}
\caption{
The energy dependence of  mean ($M_{p-\bar p}$) (top, left),
square root of the variance ($\sigma_{p-\bar p}$) (top, right),
skewness ($S_{p-\bar p}$) (bottom, left) and kurtosis ($K_{p-\bar p}$)
(bottom, right) of net proton number fluctuations.
The points are the RHIC results obtained by the STAR collaboration \protect\cite{STAR}.
The lines are  calculated in the rescaled (see text) HRG model on the freeze out curve.
}
\label{fig:mnl}
\end{center}
\end{figure}

The change of volume  with energy on the freeze out line is calculated by comparing 
data on $dN_{\pi^-}/dy$ at mid rapidity for different $\sqrt s$  with HRG model 
results.\footnote{For a compilation of heavy ion data on charged pion yields at 
mid--rapidity at different energies see e.g. Ref. \cite{andro1}.} We then obtain,
$V=[dN_{\pi^-}/dy]^{data}/n_{\pi^-}^{HRG}[T,\vec\mu]$, where in the HRG model the 
negatively charged pion density $n_{\pi^-}^{HRG}$  is calculated using the relation
between $\sqrt s$ and the thermal parameters given in Eqs.~\ref{TmuB} and \ref{muB}. 
Our results  on $V(\sqrt s)$ extracted in this way  are consistent with those obtained 
recently in Ref. \cite{andronic}.

Fig.~\ref{fig:mnl} (top left) shows the energy dependence of the first moment
($M_{p-\bar p}$) of  net proton number in the HRG model with a volume
parameter, $VT^3$, rescaled by the factor $K\simeq 0.22$. One can see in this figure 
that the HRG model results are consistent with the data.

Taking into account the results for various ratios of moments shown in Fig.~\ref{skew}
it immediately follows that the HRG model will also describe the energy dependence
of other moments, {\it i.e.} variance, skewness and kurtosis. These are also
shown in Fig.~\ref{fig:mnl}.

The good agreement of HRG model calculations with RHIC data allows us
to make predictions for different moments of charge fluctuations
covering the energy range of the RHIC low energy scan and the lowest energy
for heavy ion collision at the LHC.
We  summarize the HRG model results at different  energies in Table~\ref{tab:future}.

\subsection{Electric charge and strangeness fluctuations}

\begin{figure}
\begin{center}
\epsfig{file=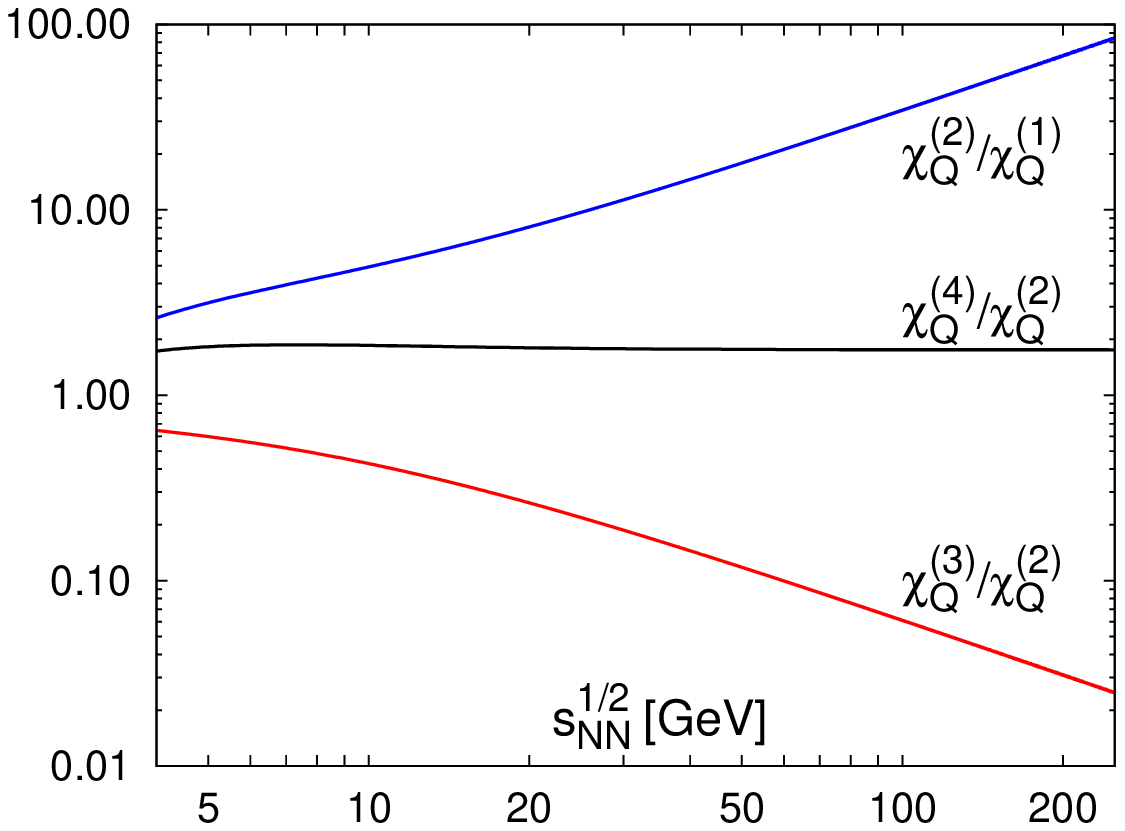,width=6.9cm}
\hspace*{-0.5cm}
\epsfig{file=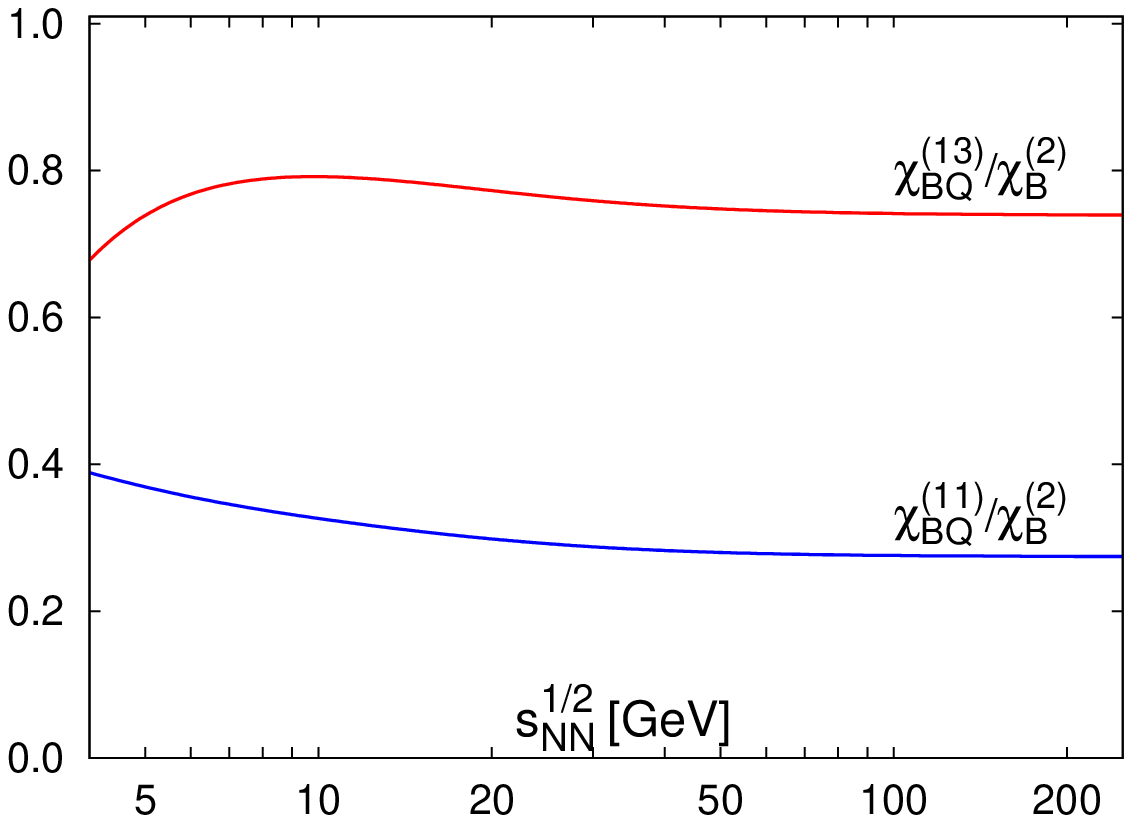,width=6.9cm}
\end{center}
\caption{\label{fig:SQ}
The ratio of moments of electric charge
fluctuations on the freeze-out curve (left) and correlations between
baryon number and electric charge fluctuations (right).}
\end{figure}

More subtle dependencies on temperature and baryon number arise
in the case of electric charge and strangeness fluctuations where multiple
charged hadrons  get larger weight in higher moments
and where meson as well as baryon sectors contribute.
This results in characteristic deviations of the kurtosis, more precisely
$\kappa_Q\sigma_Q^2=\chi_Q^{(4)}/\chi_Q^{(2)}$, from unity and also the
skewness no longer is simply related to $\tanh(\mu_B/T)$.
In the case of electric charge fluctuations we may separate contributions
of different charge sectors to the partition function. For instance, for $n$ even, we
may then obtain for moments of electric
charge fluctuations,

\begin{equation}
\chi^{(n)}_Q = \frac{1}{VT^3} \left( \ln Z_{|Q|=1}(T,\mu_B,\mu_Q,\mu_S)
+2^n \ln Z_{|Q|=2}(T,\mu_B,\mu_Q,\mu_S) \right) \ ,
\label{HRG_Q_fluctuations}
\end{equation}
where the logarithms of partition functions, $\ln Z_{|Q|}$, are obtained from
Eq.~\ref{eqq1} by
restricting the sums over mesons and baryons to the relevant charge sectors.
From this it is obvious that deviations of $\kappa_Q\sigma^2_Q=\chi_Q^{(4)}/\chi_Q^{(2)}$
from unity only arises from contributions of baryons with electric charge 2.
Similarly the odd moments are modified.
On the freeze-out curve this leads to a characteristic dependence of
ratios of moments on the collision energy that is shown in Fig.~\ref{fig:SQ}.
In the energy range relevant for current low-energy runs at RHIC \cite{lowE}
as well as at LHC  one has $\kappa_Q\sigma^2_Q\simeq 1.8$, which varies only
little with $\sqrt{s_{NN}}$.

In addition one may analyze correlations between baryon number and different
moments of charge fluctuations. Some results  are shown in the left hand
part of Fig.~\ref{fig:SQ}.

\begin{figure}
\begin{center}
\epsfig{file=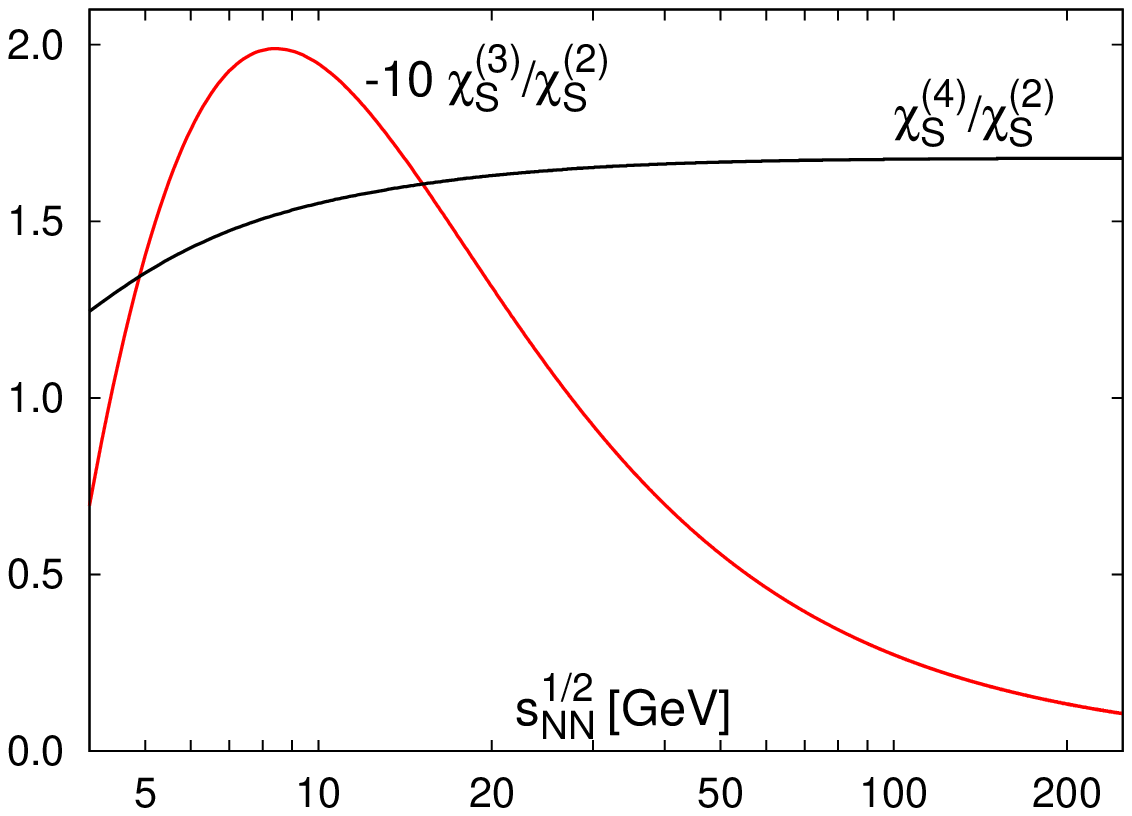,width=6.9cm}\hspace*{-0.5cm}
\epsfig{file=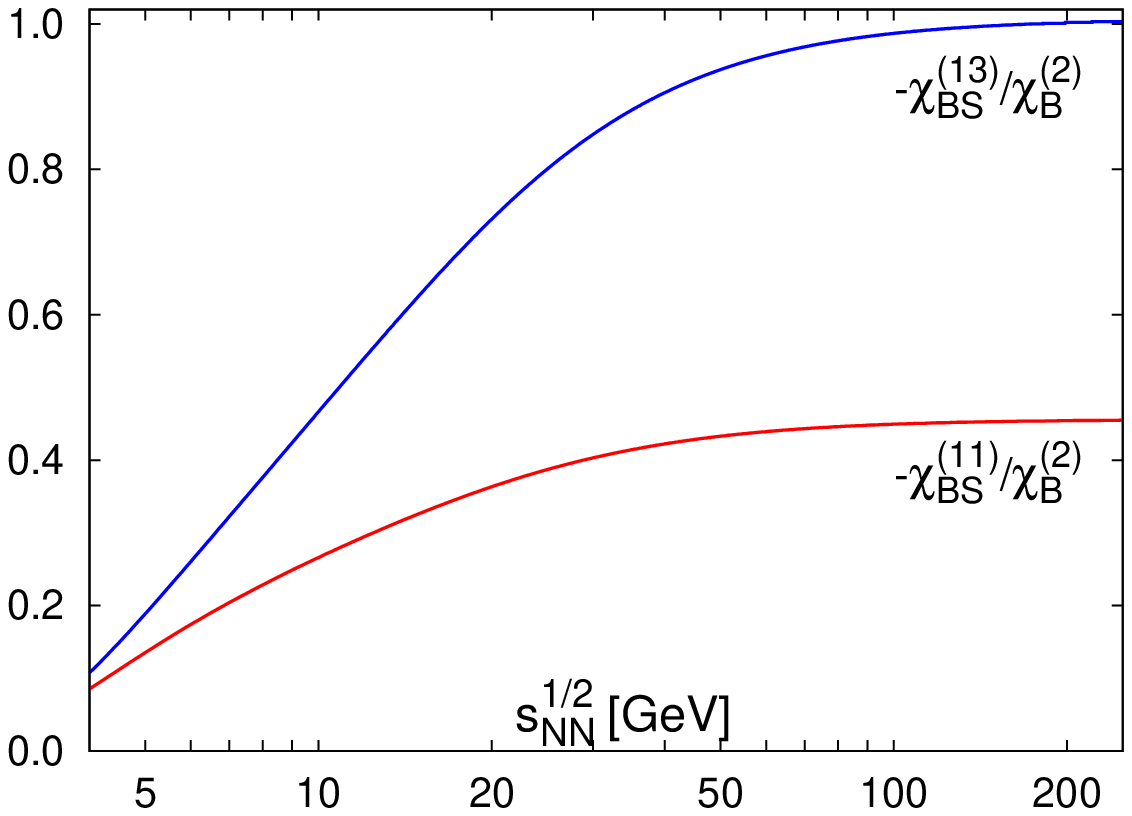,width=6.9cm}
\end{center}
\caption{\label{fig:BQS}
The ratio of moments of strangeness
fluctuations on the freeze-out curve (left) and correlations between
baryon number and strangeness fluctuations (right).
}
\end{figure}

For completeness we show in Fig.~\ref{fig:BQS} fluctuations and correlations
in the strange\-ness sector of the HRG model. In practice it may be more
difficult to compare this with experimental results as it will be crucial
that the analysis allows for strangeness fluctuations in a sub-volume and
will not impose the constraint of vanishing strangeness on event-by-event
basis.

\section{Lattice QCD and the Hadron Resonance Gas}

The above considerations suggest that results on quadratic, cubic and
quartic fluctuations of baryon number are in good agreement with
HRG model predictions on the freeze-out curve at least for collision
energies ranging from $\sqrt{s_{NN}}=200$, 62.4  down to $19.4$~GeV.
In terms of thermodynamic parameters the currently analyzed range of
collision energies at RHIC covers a rather narrow temperature
regime, $160~{\rm MeV}\le T\le 165~{\rm MeV}$. Changes in  different moments of
charge fluctuations  with collision energy are  mainly  due to the variation of  the
baryon chemical potential along the freeze-out curve which covers the range
$23~{\rm MeV} \le \mu_B \le 210~{\rm MeV}$.

To the extent that the freeze-out temperature for $\mu_B/T\simeq 0$ seems to be
close to the transition temperature from hadronic matter to the plasma phase,
the experiments cover a regime  where $0.96 \lsim T/T_c \lsim 1.0$. In this narrow
temperature range and for quark chemical potentials that are significantly smaller
than $T$, {\it i.e.} $\mu_q/T = \mu_B/3T \lsim 0.44$, there exists a lot of information
on charge fluctuations from lattice QCD calculations. Due to the rather small value
of $\mu_q/T$ already lowest order Taylor expansions of quark number susceptibilities
provide good guidance for the behavior of susceptibilities on the freeze-out curve
\cite{Gavai,Taylor,Schmidt}.
Ratios of susceptibilities have been shown to be consistent with HRG model calculations
even close to the transition temperature. To this extent the results obtained at RHIC
are not only consistent with HRG model calculations but are also consistent
with lattice QCD results.

It may be conceivable that hadron fluctuations will only
be sensitive to thermodynamics at freeze-out. One thus may ask whether fluctuations
of a completely thermalized system at chemical freeze-out in heavy ion collisions,
bare any knowledge about a nearby critical point in the QCD phase diagram.
In the following we will argue, that this is indeed the case.

In fact, we know for sure from lattice calculations that at vanishing
baryon chemical potential and for temperatures close to the transition
temperature the HRG model has to fail in describing thermal
moments of sufficiently high order. This reflects the existence of a
chiral phase transition in QCD at vanishing light quark mass. It is
a consequence of chiral symmetry restoration that moments $\chi_B^{(n)}$
will diverge at $T_c(\mu_B=0)$ for $n\ge 6$. At  non- vanishing
quark mass this leads to an oscillatory behavior of moments
close to $T_c$; e.g. $\chi_B^{(6)}$ will vanish at the transition temperature
\cite{Taylor,Schmidt}. A direct consequence of the analytic structure of
the QCD partition function in the transition region is that
$\chi_B^{(8)}$ will become negative. This has been observed in lattice
calculations \cite{RBC} as well as in chiral models \cite{chiral}.
In the HRG model, on the other hand, all  moments of  baryon
number fluctuations are positive. In fact, the ratio $\chi_B^{(6)}/\chi_B^{(2)}$
will be unity in the HRG model for the same reason as $\kappa_B\sigma_B^2=1$.
In particular at $\mu_B\simeq 0$, where the freeze-out curve is expected to
be closest to the QCD transition curve or may even coincide with it,
one should find strong deviations from unity. Lattice calculations suggest that
$\chi_B^{(6)}/\chi_B^{(2)}$ vanishes at the pseudo-critical temperatures and
rapidly rises for temperatures below, but close, to the transition temperature
\cite{Schmidt}. An
analysis of these high order moments at LHC energies clearly would have the
best chance to reveal differences between HRG model and lattice QCD
calculations and to find evidence for critical behavior 
already at $\mu_B/T\simeq 0$.

Current results on $\chi_B^{(6)}/\chi_B^{(2)}$ are still noisy \cite{Schmidt} but
suggest that also this quantity may be well described by
the HRG model up to temperatures close to the transition
temperature. Such an agreement will stop to hold closer to the transition 
temperature. However,
a better determination of this quantity in the transition region
is necessary to quantify this and to confront it with experimental data. 
Eventually, it may be necessary to evaluate also
$\chi_B^{(8)}/\chi_B^{(2)}$ ratio in order to observe striking
differences between HRG model calculations and lattice QCD results
in the energy range currently covered by RHIC and LHC experiments.

Of course, if the critical point exist in QCD and if the hadronic freeze-out occurs
within the critical region, then already the second moment of baryon number and
electric charge fluctuations should  deviate from the HRG model result. The higher
order cumulants  should   exhibit even stronger sensitivity to critical fluctuations
showing larger deviations from the model predictions. New insight into this is
to be expected to come from the first low energy run at RHIC which has been completed
this year.

\begin{table}
\begin{center}
\begin{tabular}{|c|c|c|c|c|c|c|}
\hline
$\sqrt{s_{NN}}$ & $\chi_{B}^{(2)}/\chi_B^{(1)}$ & $\chi_{B}^{(3)}/\chi_B^{(2)}$ &
$\chi_{Q}^{(2)}/\chi_Q^{(1)}$ & $\chi_{Q}^{(3)}/\chi_Q^{(2)}$ &
$\chi_{BQ}^{(11)}/\chi_B^{(2)}$ & $\chi_{BQ}^{(31)}/\chi_B^{(2)}$  \\[2pt]
\hline
7.7  & 1.01 & 0.99 & 4.18 & 0.49 & 0.34 & 0.79 \\
11.5 & 1.05 & 0.95 & 5.39 & 0.39 & 0.32 & 0.79 \\
19.6 & 1.23 & 0.81 & 7.95 & 0.27 & 0.30 & 0.77 \\
39.0 & 1.87 & 0.53 &14.25 & 0.15 & 0.28 & 0.75 \\
62.4 & 2.75 & 0.36 &21.97 & 0.09 & 0.28 & 0.74 \\
200.0 &8.20 & 0.12 &67.80 & 0.03 & 0.27 & 0.74 \\
\hline
2760 &111.1 &0.09 &922.4 &0.02 &0.27 & 0.74 \\
\hline
\end{tabular}
\caption{Ratios of moments of baryon number and electric charge fluctuations
as well as their correlations for several values of the collision energy
(in units of GeV) covering
the RHIC low energy run and LHC (last row) energies. Furthermore, one has
$\kappa_B\sigma_B^2=\chi_B^{(4)}/\chi_B^{(2)}=1$ in the entire energy range and
$\kappa_Q\sigma_Q^2=\chi_Q^{(4)}/\chi_Q^{(2)}$ varies from 1.85 at low energies
to 1.75 at high energies.}
\label{tab:future}
\end{center}
\end{table}

\section{Conclusion}

We have discussed  properties of net charge fluctuations in nuclear  matter within the
hadron resonance gas (HRG) model. We have focused  on the behavior of fluctuations
related to  baryon number, strangeness and electric charge conservation. Based on
the phenomenological relation between thermal parameters and collision energy in heavy
ion collisions we have calculated ratios of different moments of these fluctuations
along the chemical freeze-out curve where secondary hadrons exhibit thermal and chemical
equilibrium. We have also considered the energy dependence of  baryon-strangeness and
baryon-charge correlations in the HRG model. Our calculations  covered the energy
range of   $\sqrt s_{NN} > 4$ GeV  where the description of fluctuations within  the HRG
model formulated in the  grand canonical ensemble  is adequate.

Establishing  generic properties of  fluctuations in the HRG model we have compared its
predictions with the recent data of the STAR Collaboration at RHIC  on baryon number
fluctuations, expressed by  net proton fluctuations, determined in Au-Au collisions at
three different collision energies. We have shown, that the STAR data
are consistent with  the HRG model results.  The change of measured fluctuations with
collision energy can be accounted for by the variation of the baryon chemical
potential along the freeze-out curve.
A list of
results from HRG model calculations at collision energies that include the
published STAR results, the parameters of the 2010 low energy scan at RHIC and
collision energies at the LHC are given in Table~\ref{tab:future}.

We have argued that the apparent agreement of the HRG model with the STAR data does not
necessarily mean that in heavy ion collisions at chemical freeze-out the system has
lost entirely its  memory of the expansion period during which it
may have passed through a region of the QCD phase diagram where correlation lengths
are large, as expected in the vicinity of a critical point. If freeze-out
occurs close to or in the transition region between hadronic matter and
quark-gluon plasma this will, even in the absence of a phase transition,
show up in higher order moments. The first four moments of baryon number density
fluctuations agree well between HRG model and  lattice QCD calculations even
close to $T_c$.  Deviations
form the HRG model results, however, have been seen  in higher order moments. We have
argued  that in order to observe critical fluctuations in heavy ion collisions  one
would possibly need to measure even higher order moments.

Our results on energy dependence of different moments  and their specific relations
in the HRG model can  be used  to characterize 'ordinary' thermal properties
of higher order moments of charge fluctuations and to set a baseline for the
observation of  critical fluctuations in
nuclear matter created in heavy ion collisions.

\section*{Acknowledgments}
\addcontentsline{toc}{section}{Acknowledgements}

We  acknowledge  stimulating discussions with  Tapan Nayak and Nu Xu.
KR also acknowledges fruitful discussions with A. Andronic and B. Friman and
the partial support of the Polish Ministry of Science.
This work has been supported in part by contract DE-AC02-98CH10886 with the
U.S. Department of Energy.

\end{document}